\documentclass[11pt,times]{article}
\usepackage[left=1in,top=0.9in,right=1in,bottom=1in]{geometry}
\usepackage[numbers]{natbib}
\usepackage[small,compact]{titlesec}
\usepackage{titling}
\posttitle{\par\end{center}}
\usepackage{authblk}
\title{{\large \textbf{The impact of astrophysical particle acceleration on searches for beyond-the-Standard-Model physics}}
}
\author[1]{\small A. Weinstein}
\author[2]{\small J. Dumm}
\author[2]{\small L. Fortson}
\author[3]{\small R. Mukherjee}
\affil[1]{\emph{\small Iowa State University}}
\affil[2]{\emph{\small University of Minnesota}}
\affil[3]{\emph{\small Barnard College, Columbia University}}
\date{}

\begin{document}
\maketitle

\vskip -0.5in

Signatures of physics beyond the Standard Model can be imprinted in the spectrum of both charged particles and photon radiation produced by either individual astrophysical sources or astrophysical source populations.  Searches using astrophysical probes are often an essential complement to direct detection and accelerator experiments, since they can probe otherwise inaccessible portions of new model (e.g supersymmetric, GUT) parameter space.  Unequivocal interpretation of these results will require a good understanding of the particle acceleration mechanisms taking place within astrophysical sources, as these acceleration processes may mimic or mask signatures of new physics.  This may occur due to effects within a particular source being used as a probe (as in tests of Lorentz invariance violation or searches for axion-like particles), an unexpected contribution from a class of astrophysical sources (as with measurements of the positron spectrum) or the presence of a large background from other astrophysical sources that can swamp a weak signal (indirect dark matter (DM) searches).

\section*{Impact on Indirect Searches for Dark Matter}

\textbf{The Positron Spectrum:} The positron spectrum measured by satellite and balloon-borne experiments can provide an indirect probe of DM, since in some models, WIMP DM may annihilate in the galactic halo to produce positrons.  As the known contribution to the local positron spectrum is from particles accelerated within Galactic astrophysical sources, successful interpretation of any anomaly in the positron spectrum clearly depends on our understanding of the acceleration processes occurring within distinct populations of astrophysical sources: pulsars, pulsar wind nebulae (PWNe) and supernova remnants SNRs).  Current discussions of the results from PAMELA\citep{2009Natur.458..607A}, Fermi\citep{2009PhRvL.102r1101A} and AMS\citep{PhysRevLett.110.141102}, which show that the positron fraction increases with energy in a way not predicted by the usual models of the astrophysical background, clearly illustrate this principle.  While the PAMELA, Fermi, and AMS results can be accommodated by annihilating DM, astrophysical sources also produce electron-positron pairs.  It has been suggested the positron excess could arise because diffuse shock acceleration hardens the energy spectrum of secondary positrons produced in SNRs\cite{2009PhRvD..80l3017A,2009PhRvL.103e1104B}.
Electrons can also be accelerated in the pulsar magnetosphere (with the precise region depending on the emission model) and low energy electrons can escape along open field lines or as part of the pulsar wind\cite{2009JCAP...01..025H}.  It is thought that the chief contributors to the positron spectrum in this case would be mature pulsars, since electrons and positrons produced by young pulsars are likely to be confined by the associated PWN.  It has been shown that, given current knowledge, mature pulsars provide a viable alternative explanation of positron fraction excess \citep{2009JCAP...01..025H,2009PhLB..678..283B}.  However, these predictions depend critically on the assumption that magnetic dipole radiation dominates the pulsar's energy loss and are sensitive to details of the pulsar population, from individual pulsar spectra to the distribution and ages of pulsars throughout the Galaxy\citep{2009JCAP...01..025H}.

\noindent\textbf{Searches for Dark Matter Annihilation:} A number of beyond-the-Standard-Model scenarios, particularly certain classes of supersymmetric models, predict a new WIMP (weakly interacting massive particle) that is a viable candidate for DM.  For the classes of models for which this WIMP is its own antiparticle, DM could reveal itself via self-annihilation to the $\gamma\gamma$, $\gamma Z$, and continuum channels in regions of high DM density.  Detection of a gamma-ray signal from DM annihilation would provide an unequivocal indication of the nature and mass of astrophysical DM\citep{Doro2013189}.  Observational targets for indirect dark matter searches with gamma rays include any region of enhanced dark matter density, including our own Galactic Center (GC), dwarf satellite galaxies, subhalos in the Galactic halo, and galaxy clusters\citep{Doro2013189,2011JCAP...12..011S} (see also whitepaper \cite{2013arXiv1305.0302W}).

DM annihilation signals from the GC, subhalos in the Galactic halo, or galaxy clusters must be disentangled from the gamma-ray emission produced by particle acceleration processes.  The problem of distinguishing DM annihilation emission from subhalos from other astrophysical sources is relatively straightforward (see the whitepaper \cite{2013arXiv1305.0312N} for further details.)  In the case of the GC and galaxy clusters, the problem is more difficult.  Analysis of the spectrum of the strong gamma-ray point source at the Galactic Center, HESS J1745-290, shows that it is dominated by non-thermal emission of astrophysical origin\citep{2004A&A...425L..13A,2006PhRvL..97v1102A}.  A DM annihilation component to this emission must be less than $10\%$ of the point source flux and must also be disentangled from diffuse emission thought to originate from cosmic ray interactions with giant molecular clouds\cite{2006Natur.439..695A}.

Several types of astrophysical sources contribute to gamma-ray emission in galaxy clusters that can mask a DM annihilation signal.  Clusters may contain bright Active Galactic Nuclei (AGN) that may interfere with possible DM detection.  In the GeV-TeV range accessible to IACTs, the probability of intense AGN emission is somewhat reduced due to decreasing inverse Compton (IC) scattering efficiency, but many AGN are still known to emit efficiently at very high energies.  Cosmic rays (CRs) also constitute a likely source of background to DM searches with galaxy clusters.  Both CR electrons and protons can be injected into the intra-cluster medium by structure formation shocks, radio galaxies and supernovae-driven galactic winds \citep{2010ApJ...710..634A,2011JCAP...12..011S}.  CR electrons reveal themselves in many galaxy clusters via synchrotron radiation in radio (radio cluster halos)\citep{2008SSRv..134..191P,2008SSRv..134...93F} and may produce GeV-TeV emission via IC up-scattering of the cosmic microwave background\citep{2002MNRAS.337..199M,2010ApJ...710..634A, 2011JCAP...12..011S}.  The dominant component of the non-DM gamma-ray emission from galaxy clusters is expected to come from the decay of neutral pions, produced by the CR protons interacting with the intracluster medium (ICM)\citep{1996SSRv...75..279V,Aleksic:2010eg}.


\textbf{Searches for Axion-like Particles:} Light spin-zero bosons with a two-photon coupling, dubbed axion-like particles (ALPs), arise as a consequence of many grand unified theory (GUT) models.  Like axions, some ALPs are potential cold dark matter candidates as well as clear signatures of physics beyond the Standard Model; unlike axions, a broad range of ALP parameter space gives rise to measurable astrophysical effects due to photon-ALP mixing in the presence of both intergalactic magnetic fields (IGMFs) and magnetic fields within astrophysical sources.  These effects would be imprinted on the spectra of cosmological VHE gamma-ray sources such as AGN; moreover,  mixing due to IGMFs could suppress the absorption feature expected in AGN spectra due to pair-production off of the extragalactic background light (EBL)\citep{Horns:2012ct}. (See the whitepaper \cite{2013arXiv1305.0252S} for further details). However, standard particle acceleration processes within AGN could mimic this suppression effect.  In particular, if AGN are the origin of ultra-high-energy cosmic rays (UHECR), these cosmic rays may interact with the EBL to produce TeV photons via both photon pair production and pion production.  It is expected that these secondary gamma-rays will appear nearly point-like for current instruments and could modify the observed gamma-ray spectrum of AGN at high energies\citep{2011ApJ...731...51E}.
For several blazar spectra, the size and nature of this effect has been shown to be largely independent of the EBL and cosmic ray injection spectrum, depending only on the overall power emitted by the source in cosmic rays \cite{2011ApJ...731...51E}.  New higher quality spectra from existing IACTs or CTA may permit us to probe small differences in how AGN spectra are modified by the range of allowed EBL models\cite{2011ApJ...731...51E}.  (See whitepapers \cite{2013arXiv1305.0253D} and \cite{2013arXiv1304.8057K} for further details).

\section*{Impact on Tests of Lorentz Invariance}

Modifications of the vacuum dispersion relation due to Lorentz invariance violation (LIV) can result in a measurable propagation delay between photons of different energies from an astrophysical source.  This effect will be measurable when the emission from an astrophysical source has a clearly identifiable structure in the time domain (such as sharp transients or emission with a periodic structure, as from pulsars).  (See the whitepaper \cite{2013arXiv1305.0264O} for further details).
Constraints on LIV have been obtained through studies of gamma-ray bursts (GRBs), flares from active galactic nuclei (AGN), and pulsars.  Interpretation of LIV measurements made using these source must address the degree to which processes within the source must mimic LIV.  This is particularly challenging in the case of GRBs, since the particle aceeleration processes that lead to GRB gamma-ray emission are largely unknown.

\section*{Role of the Cherenkov Telescope Array}

The Cherenkov Telescope Array (CTA) will significantly improve both our understanding of the particle acceleration processes taking places in particular astrophysical source classes and our ability to distinguish effects of these particle acceleration processes from signatures of new physics.  For example, CTA will noticeably improve our ability to do population studies with astrophysical probes such as AGN.  CTA's order of magnitude leap in sensitivity, combined with coverage of both Northern and Southern hemispheres, will dramatically increase the number of AGN detections; CTA's greater low-energy reach (where absorption effects due to the EBL are negligible) will extend detections out to greater redshifts, possibly as far as $z \sim 2$ \citep{Sol2013215}.  Estimates extrapolated from the Fermi catalog of AGN detected below $\sim 100$ GeV suggest a gain of nearly 200 sources in less than three years of operation.  CTA's increased effective area also translates to an ability to detect variability on shorter timescales\citep{Sol2013215}; since the secondary gamma ray emission from UHECR (which could mimic an ALP signature) is not expected to be variable, this could be the key to differentiating between UHECR and ALP modifications to AGN spectra.  The ability to detect variability from AGN and other sources on short timescales is also critical for LIV studies.

Moreover, CTA's reduced point spread function (PSF) with respect to earlier gamma-ray telescopes allows easier removal of AGN foregrounds in DM searches using galaxy clusters, better resolution of different point source and diffuse emission components from the GC, and could allow for detection of the slight smearing of the UHECR secondary gamma rays from AGN.  CTA's extremely wide field of view ($8^{\circ}$ as opposed to the $3-5^{\circ}$ of current-generation Cherenkov telescopes) makes it possible to detect highly extended ($>1^{\circ}-2^{\circ}$) gamma-ray sources\citep{Hinton20131,Acharya20133}.  Since both the DM annihilation and CR-induced gamma-ray emission from galaxy clusters is expected to be highly extended, but on different angular scales\citep{2011JCAP...12..011S}, this could make a substantial difference to detecting CR-induced gamma-ray emission and disentangling it from a possible DM signature.

In all cases, the US contribution of mid-sized telescopes (MSTs) would be particularly important, as it improves the sensitivity of the core energy regime (100 GeV - 10 TeV) of CTA by a factor of two to three \cite{Jogler:2012tc,Hinton20131,Acharya20133,2012SPIE.8444E..18C}.  Moreover, the high pixelation of the US MST camera permits this subarray to improve the gamma-ray PSF relative to the baseline CTA MSTs by as much as $40\%$ at certain energies.

\small

\bibliographystyle{plain}

\end{document}